\documentclass[12pt,preprint]{aastex}
\usepackage{natbib}
\usepackage{graphicx}

\input epsf

\begin{document}

\title{Massive Clumps in Local Galaxies: Comparisons with High-Redshift Clumps}

\author{
Bruce G. Elmegreen\altaffilmark{1}, Debra Meloy Elmegreen\altaffilmark{2}, J.
S\'anchez Almeida\altaffilmark{3}, C. Mu\~noz-Tu\~n\'on\altaffilmark{3}, J.
Dewberry\altaffilmark{2}, J. Putko\altaffilmark{2,4}, Y. Teich\altaffilmark{2},
and M. Popinchalk\altaffilmark{2,5} }

\altaffiltext{1}{IBM Research Division, T.J. Watson Research Center, Yorktown Hts.,
NY 10598} \altaffiltext{2}{Vassar College, Dept. of Physics and Astronomy,
Poughkeepsie, NY 12604} \altaffiltext{3}{Instituto de Astrof'sica de Canarias, C/
via L\'actea, s/n, 38205, La Laguna, Tenerife, Spain} \altaffiltext{4}{Middlebury
College, Dept. of Physics, Middlebury, VT 05753} \altaffiltext{5}{Wesleyan
University, Dept. of Astronomy, Middletown, CT 06459}
\begin{abstract}
Local UV-bright galaxies in the Kiso survey include clumpy systems with kpc-size
star complexes that resemble clumpy young galaxies in surveys at high redshift. We
compare clump masses and underlying disks in several dozen galaxies from each of
these surveys to the star complexes and disks of normal spirals. Photometry and
spectroscopy for the Kiso and spiral sample come from the Sloan Digital Sky
Survey. We find that the largest Kiso clumpy galaxies resemble UDF clumpies in
terms of the star formation rates, clump masses, and clump surface densities.
Clump masses and surface densities in normal spirals are smaller.  If the clump
masses are proportional to the turbulent Jeans mass in the interstellar medium,
then for the most luminous galaxies in the sequence of normal:Kiso:UDF, the
turbulent speeds and surface densities increase in the proportions 1.0:4.7:5.0 and
1.0:4.0:5.1, respectively, for fixed restframe B-band absolute magnitude. For the
least luminous galaxies in the overlapping magnitude range, the turbulent speed
and surface density trends are 1.0:2.7:7.4 and 1.0:1.4:3.0, respectively. We also
find that while all three types have radially decreasing disk intensities when
measured with ellipse-fit azimuthal averages, the average profiles are more
irregular for UDF clumpies (which are viewed in their restframe UV) than for Kiso
galaxies (viewed at g-band), and major axis intensity scans are even more
irregular for the UDF than Kiso galaxies. Local clumpy galaxies in the Kiso survey
appear to be intermediate between UDF clumpies and normal spirals.
\end{abstract}


\section{Introduction}

Disk galaxies at increasing redshift become more irregular \citep{abraham96,
conselice05} and clumpy (Elmegreen et al. 2004; see review in Shapley 2011) as a
result of an increased frequency of mergers in the denser universe
\citep{kartaltepe12,puech12,kaviraj13,mclure13}, and because of higher gas fractions
\citep{tacconi13,bothwell13} and velocity dispersions \citep{law09,for09,newman13}
that lead to more violent gravitational instabilities
\citep{ee05,bournaud07,genzel08,ceverino10,wisnioski11}. Over time, mergers become
less frequent, and an increasing stellar mass fraction gives a galaxy more
stability, with density waves in the stars replacing violent clump formation in the
gas \citep{cacciato12}.

Local galaxies with relatively large star-forming complexes are nearby analogs of
high redshift clumpy galaxies that allow a comparison of large-scale star forming
regions over time. For example, local dwarf irregulars resemble high redshift
galaxies with respect to gas fraction, relative clump size, and relative thickness
\citep{elmegreen09a}. Local tadpole galaxies \citep{etadpole12}, which include the
sub-class iI,c \citep{noeske00} of Blue Compact Dwarfs \citep[BCDs, see
also][]{paz,kniazev} and a high fraction of extremely metal-poor BCDs
\citep{pap,morales}, resemble high redshift tadpoles, which account for 10\% of
large galaxies in the Hubble Ultra Deep Field \citep{elmegreen07,etadpole12}.
Tadpoles contain one giant clump of stellar mass $M_{\rm *}\sim10^7\;M_\odot$
\citep{etadpole12} with a tail to the side. The clumps in several local tadpoles
have lower metallicities than the rest of the galaxy, suggestive of accretion
\citep{sanchezalmeida13}.  Similar localized metallicity drops are observed in
$M_{\rm *}\sim10^{10}\;M_\odot$ star-forming galaxies at high redshift
\cite[e.g.,][]{cresci10,queyrel12}. Some BCDs suggest active accretion as well,
considering the irregular and sometimes non-rotating pools of HI gas that surround
them \citep{brinks88,taylor96, thuan97b,
vanzee98a,vanzee98b,putman98,wilcots98,pustilnik01,hoffman03},

Luminous Blue Compact Galaxies are another type of local analog that are small,
clumpy and irregular with high gas fractions ($\gtrsim10$\%) and distorted
kinematics \citep{garland07}. They are locally rare, possibly interacting
\citep{garland04}, and like intermediate-redshift ($z\sim0.5$) blue compact
galaxies \citep{hoyos07}, they may evolve into dwarf ellipticals or low-mass
irregulars with dynamical masses of $10^{10}-10^{11}\;M_\odot$.

Other local analogs of high redshift galaxies have higher masses than BCDs or
Luminous Blue Compact Galaxies.  \cite{casini76} studied the massive clumpy
irregular galaxies Markarian 7, 8, 296, and 325, in a survey of Markarian galaxy
pairs. With considerable foresight, they suggested that the clumps could be galaxy
fragments and that the galaxies might settle down to normal Hubble types over
time. \cite{heidmann87} summarized this rare type of galaxy noting that the clumps
have 100 times the luminosity of 30 Doradus \citep{benvenuti82} and
$\sim10^8\;M_\odot$ of ionized gas; Mrk 325 has x-ray emission equivalent to
$10^4$ supernovae remnants. In Mrk 296, there are several large clumps along a
jagged line, making it look like a crooked chain galaxy; the share of the total
stellar mass in each clump is $\sim10^8\;M_\odot$ \citep{casini79}. Three other
massive clumpy galaxies were subsequently studied from the Kiso Ultraviolet Galaxy
survey by \cite{maehara88}: 1618+378, 1624+404, 1626+413, to which a fourth was
added, Mrk 297.  In terms of clump mass, these galaxies resemble high-redshift
clumpy galaxies. \cite{maehara88} state that they are $\sim10$ times brighter and
$\sim3$ times larger than classic irregulars, and the most distant ones are at 120
Mpc.

Local Ultraviolet Luminous Galaxies also resemble star-forming galaxies at $z>1$
\citep{heckman05}.  The most compact of these are brighter in the UV than local
BCDs, having $M_{\rm *}\sim10^{10}\;M_\odot$, and resemble Lyman Break Galaxies in
luminosity, mass, metallicity and star formation rate \citep{hoopes07}.  Lyman
Break Analogs like this also compare well morphologically to clumpy high redshift
galaxies \citep{overzier08,overzier09,overzier10}. Similar studies by
\cite{hibbard97}, \cite{barden08}, \cite{petty09} and others show significant loss
of faint, small, and peripheral structures in local starburst galaxies when viewed
as if they were at high redshift, so that only the most clumpy and irregular of
the nearby systems, many of which are interacting, resemble young galaxies.
\cite{petty09} suggest that Mrk 8, studied also by \cite{casini76}, NGC 3079, and
NGC 7673 resemble clumpy star-forming galaxies in the $z=1.5$ to 4 range.

Here we study low-redshift clumpy galaxies from the Kiso Survey for
Ultraviolet-Excess Galaxies \citep{kiso-all}.  This is a survey of $\sim10^4$
galaxies brighter than 18th magnitude (photographic) covering 7000 square degrees
made with the Kiso Schmidt telescope. The survey classified uv-bright galaxies
according to various morphologies using the Palomar Observatory Sky Survey and the
Sloan Digital Sky Survey \citep[SDSS;][]{stoughton}. We are interested in
classification type Ic, which has large blue clumps, and type Ig, which has a
single large clump. The survey contains 176 Ic and 83 Ig galaxies, of which 158 of
the 259 total are in the SDSS.  Many of these 158 galaxies are spirals with one or
more bright star-forming regions. Others are irregulars or possible interactions;
95 have SDSS-DR7 spectra \citep{aba}. A sample of 36 of those with spectra were
chosen for the present study, based on their low inclination, relative isolation,
and clumpiness.

In Section 2, we measure the magnitudes in five SDSS filters of each prominent
clump in the Kiso sample and determine the masses, ages and extinctions in these
clumps from population synthesis models. In Section 3, we also determine the
galaxy star formation rates from the SDSS single-fiber spectra. The Kiso clumps
are compared with star-forming clumps in 20 representative local spiral galaxies
covering a range of Hubble types and arm classes. These results are then compared
with the properties of clumps in high-redshift clumpy galaxies of the
``clump-cluster'' type \citep{elmegreen05a}.  Radial light profiles are also
compared in the different galaxy samples in Section 4. A discussion is in Section
5, and conclusions are in Section 6.

\section{Star-forming Clumps}
\subsection{Photometry and Modeling}

Archival images of 36 clumpy UV-bright galaxies in the Kiso survey were obtained
in $u,g,r,i,z$ filters from the SDSS survey website (www.sdss.org), as listed in
Table 1 and shown in Figures 1-3. The Kiso Ultraviolet Galaxy (KUG) numbers in
Table 1 come from the Kiso2000 catalog \citep{kiso-all}; we also give the Kiso
number, which is the catalog entry number minus 1. Although the galaxies in this
sample were selected because of their bright star-forming clumps, there are many
distinct morphologies. Some, like Kiso 8491, are dominated by many large blue
clumps in an irregular pattern. Others, like Kiso 5886, have regular spiral arms
with big clumps in the arms. Kiso 2205 and 6971 are irregularly shaped with long
streamers; Kiso 5979 has a long arm that looks tidal in origin; Kiso 7186, 7256,
and 7878 have off-center bulges; Kiso 5324 and 7914 show central bars, and 7186
has an offset bar. Kiso 4170 is totally irregular. Most of the Kiso galaxies
studied here have a red central region that is displaced from the center of the
outer isophotes. These features are noted in the table, along with the Kiso Ic or
Ig type, the distance in Mpc from the NASA/IPAC Extragalactic Database (NED), and
the absolute blue magnitude (from NED).

The primary selection criterion for the measured galaxies, aside from the
requirement of a low inclination for visibility of the disks and a relative
isolation to remove obvious mergers, was the presence of large star-forming
complexes.  Examination of the images shows many other odd features, however, as
detailed in the previous paragraph. The few with off-center bulges or bars are
interesting because these asymmetries suggest a short-lived phase or an unusual
dynamical process, either of which may lead to heightened gas turbulence. General
disk irregularities suggestive of pervasive and massive gravitational instabilities
may also lead to heightened turbulence.  Such turbulence should increase the Jeans
length in the disk, making the largest star-forming complexes larger than normal. We
discuss the Jeans length and mass more in section \ref{sect:resultsm}, where we use
the observed clump properties to suggest an increase in turbulent speed.

A representative sample of 20 nearby spirals in SDSS was selected from our
previous study of arm properties \citep{elmegreen11} based on infrared images from
the Spitzer Survey of Stellar Structure in Galaxies, S$^4$G \citep{kartik}. These
galaxies are listed in Table 2 along with their Hubble type (from NED), Arm Class
\citep{ee87}, and absolute blue magnitude (from NED). They were chosen to cover
the same range of absolute B-band magnitude as the Kiso galaxies, but were not
matched to star formation rate or selected for any particular morphology. They
span Hubble types Sb through Sm and include flocculent, multiple arm, and grand
design spirals. They are a good comparison sample for examining star-forming
complexes since the Kiso galaxies are intensely starbursting and these spirals are
not.

Photometric measurements on sky-subtracted and aligned images for both the Kiso
galaxies and the local spirals were done with IRAF (Image Reduction and Analysis
Facility) to determine the masses, ages, and extinctions of the clumps. The
largest clumps were identified by eye on g-band SDSS images, and their magnitudes
were measured with circular apertures in the {\it apphot} task, which subtracts
the local background from an annulus surrounding the aperture.  The masses,
surface densities, ages, and extinctions of the clumps were estimated from this
photometry by fitting the data to models in \cite{bruzual03} that assume a
Chabrier IMF and solar metallicity (models with 0.4 times the solar metallicity
had negligible differences compared to the clump-to-clump scatter). The model
colors were obtained by integration over the low resolution spectra in
\cite{bruzual03}, weighted by the Gunn filter functions (i.e., SDSS filters). We
assume a constant star formation rate back to some starting time, which is
considered to be the region age. A range of extinctions was used for the models,
following the wavelength dependencies in \cite{calz00} and \cite{leith02}.

The average observed colors of all the Kiso clumps, averaged over the three
apertures described below, were $<u-g>=0.67\pm0.63$, $<g-r>=0.22\pm0.30$, $<r-i>=
-0.08\pm0.37$, and $<i-z>=-0.04\pm0.58$.  For the SDSS clumps, the average colors
were $0.39\pm0.54$, $0.18\pm0.33$, $-0.20\pm0.41$, and $0.06\pm0.43$, respectively.

The final values for mass, age, and extinction were determined from weighted
averages of trial models using a weighting function $e^{-0.5\epsilon\chi^2}$ for
$\chi^2$ equal to the quadratic sum of the differences between the model and
observed colors, normalized to the dispersions for each color, and for
$\epsilon=0.01$. The factor $\epsilon$ makes the weighting function smooth
considering the fact that there are many systematic uncertainties in addition to
the statistical uncertainty of noise in the measured colors. The weighted averages
were done only for the models with the lowest rms deviations from the
observations, as determined by binning the rms deviations in intervals of 0.02.
The average number of models that enter into each best-fit result is $160\pm130$.
The sum of the squares of the differences between the four observed colors and the
model colors has an rms of $0.05\pm0.06$.

Once the best-fit age and extinction are determined, the clump mass follows from
the observed $g$ magnitude in comparison to the model.  The surface densities of
star formation in the clumps are then calculated from the ratios of the masses to
the projected areas that are used to measure the magnitudes.  The telescope
resolution determines a lower limit to the measurable size of a clump and the true
sizes are smaller than the measured sizes. Using stars in several SDSS fields near
our galaxies to determine instrumental resolution, we measured an average stellar
Gaussian sigma value of about 1.2 px, depending slightly on passband.  For an
aperture size of 3 pixels, the intrinsic source size would be approximately 2.7
pixels if we subtract the resolution in quadrature. The intrinsic area would be
smaller by 16\%. Instrumental resolution does not affect the clump luminosity by a
significant amount, so the primary effect of this correction is to increase the
derived surface densities by an average factor of 1.16. In the plotted surface
densities, we do not include this increase because it is a small effect compared
to the other uncertainties, but in the determination of turbulent speed from the
surface density, we make this correction, as discussed in Section
\ref{sect:resultsm}.

For comparison with the Kiso clumps, clump properties in 30 clumpy galaxies (Table
3) in the Hubble UDF field, selected from \cite{elmegreen05a}, were determined
from Advanced Camera for Surveys (ACS) measurements in $B_{435}$, $V_{606}$,
$i_{775}$, and $z_{850}$ bands using the IRAF task {\it imstat}, which defines a
rectangle around the clump and measures the total light inside the rectangle. The
interclump background was measured from relatively clear regions nearby and
subtracted from the clump light. The same clump and interclump rectangle was used
for all passbands. The task {\it apphot} is not good for highly clumpy galaxies
because the background annulus may contain parts of other clumps. Still, a
comparison of the color measurements using {\it imstat} and {\it apphot} for 2 UDF
galaxies showed an average difference of only 0.01, 0.03, and 0.07 magnitudes for
$B-V$, $V-i$, and $i-z$, respectively.

The 30 UDF galaxies were chosen from our catalog \citep{elmegreen05a} of 178
``clump clusters'' -- so named before their kinematics
\citep{forster06,weiner06,genzel06} showed them to be whole galaxies of an
extremely clumpy type not commonly observed in the local universe.  We avoided the
more amorphous types and picked those with distinct clumps, fairly large in angle,
not obviously interacting, and not highly inclined. They span the same range of
restframe absolute B-band magnitude as the Kiso clumpy and local spiral galaxies.

The clump properties for UDF galaxies were determined following the same procedure
as for the Kiso galaxies but with redshift corrections and corrections for
intervening HI \citep{madau95}. The redshifts were photometric and taken from
\cite{rafel09}. Because high-redshift galaxies have slightly lower metallicities
for the same luminosity as local galaxies \citep[e.g.,][]{mannucci09}, we assumed
a metallicity of 0.4 solar in the \cite{bruzual03} tabulations.

As for the Kiso galaxies, telescope resolution limitations imply that the intrinsic
source size is smaller than the measured size. We measured an average Gaussian sigma
for stars in the ACS UDF field to be 3.14 px. The average {\it imstat} area for the
UDF clumps is 34.5 square pixels, and the average deconvolved area is 24.6 square
pixels. Thus the intrinsic clump areas are on average 1.4 times smaller than the
measured areas. This increases the derived surface densities by the same factor.
This factor is included in the determination of turbulent speeds below.

To compare star formation with global properties, we determined absolute restframe
B-band integrated galaxy magnitudes by interpolating the ACS filters. Restframe
B-band magnitudes were determined for the Kiso and local spiral galaxy samples by
interpolation and conversion in the SDSS filters.

We do not discuss individual clump ages below, but note that for Kiso clumps, the
average log-age (in years) is $7.1\pm1.3$ for each aperture size, and for the SDSS
galaxy clumps, it is $6.8\pm1.2$ for each aperture size.  For all of the non-bulge
UDF clumps, the average log-age is $7.8\pm0.8$. Therefore, all of the clumps have
similar ages within the dispersions. They are significantly younger than a Gyr and
typically range between 10 Myrs for local galaxies and 100 Myrs for UDF galaxies.

We also do not discuss individual extinctions below. The average fitted extinction
for all of the measured clumps in the Kiso galaxies is $1.6\pm0.5$ mag in V-band;
for the SDSS galaxies it is $1.5\pm0.6$ mag, and for the UDF galaxies it is
$2.3\pm2.0$ mag. The extinctions for UDF galaxies are fairly inaccurate without
observer-frame red measurements, but we cannot use the HST in JHK bands because
the angular resolution is too low to resolve any but the most massive clumps (it
is 3 times worse than the ACS resolution in optical bands). The masses and surface
densities discussed below have been corrected for the extinction determined by
each fit.

The derived extinctions may appear to be small for massive star-forming regions,
but the pressures from young stars should drive gas and dust away from the line of
sight, as it does in local starbursts \citep{maiz98}. Our extinctions are
comparable to those of super star clusters in the center of M82, which is the
prototype of a highly extincted nearby starburst galaxy.  The internal extinction
for 200 SSCs in the center of M82 ranges from 1.5 to 6 magnitudes in V band
\citep{melo05}. The average extinction over big sections of typical galaxies is
smaller than this. We computed the effective mean extinction in the central
regions of the Kiso galaxies from the Balmer decrements in the SDSS fiber spectra
($3^{\prime\prime}$ diameter) using the formula $A_{\rm V}=8\log({\rm
H}\alpha/(2.86{\rm H}\beta)$ \citep{dominguez13}. The average for all of the Kiso
galaxies in our survey is $A_{\rm V}=0.55\pm0.45$ mag. The largest extinction of
$A_{\rm V}=1.7$ mag corresponds to Kiso 8708. Large-scale effective mean
extinctions in spiral galaxies depend on mass but are also on the order of $A_{\rm
V}=1$ mag \citep[e.g.,][]{garn10}. These large-scale effective mean extinctions
are smaller than the extinctions to star forming regions because the large scale
dust distribution varies across a large region and the proportion of light coming
from lower-than-average extinctions outweighs the proportion coming from
higher-than-average extinctions \citep{fischera03}.

\subsection{Clump results}

\subsubsection{Mass, Surface Density, and the Jeans' Relations}
\label{sect:resultsm}

Figure \ref{measles_mass_mag} shows in the bottom three panels the non-bulge clump
masses plotted versus the absolute B-band magnitudes of the whole galaxies in all
three samples. In the top panels, the projected surface densities are shown. The
lines in each panel show the linear fits to the data (on these log-log plots). The
fit to the UDF galaxies, shown as a dotted line in the left panels, is reproduced in
the center panel for comparison with the Kiso galaxies. Similarly, the fit to the
local galaxies, a dashed line in the right-hand panels, is again in the center
panels. The solid line is the fit to the Kiso galaxies. These fits are for clump
mass $m_{\rm clump}$ and surface density $\Sigma_{\rm clump}$,
\begin{equation}
\log m_{clump}=3.64 - 0.18 M_B\;({\rm UDF});\;  -3.57-0.54 M_B\;({\rm Kiso});\;  -3.05-0.41 M_B\;({\rm Normal})
\label{eq:mass}
\end{equation}
\begin{equation}
\log \Sigma_{clump}=-1.19 - 0.12 M_B\;({\rm UDF});\;  -2.41-0.18 M_B\;({\rm Kiso});\;  -0.61-0.063 M_B\;({\rm Normal})
\label{eq:sigma}
\end{equation}

The UDF galaxies reach brighter absolute restframe B magnitudes than the Kiso
galaxies, and the most massive UDF clumps are more massive than the most massive
Kiso galaxy clumps. Also at low luminosity, the UDF clumps are more massive than
most of the Kiso clumps in galaxies of the same luminosity. However, the most
luminous Kiso galaxies have clumps that are comparable in mass and column density
to clumps in UDF galaxies at the same luminosity.  In this respect, the most
luminous Kiso galaxies are comparable to UDF clumpy galaxies of the same
luminosity. Note that both types have clumps measured in the same way and use the
same population synthesis and extinction models.

Normal local galaxies, on the other hand, have clumps that are much lower in mass
than the clumps in either the UDF or Kiso galaxies. This is to be expected because
the UDF and Kiso galaxies were chosen to be clumpy types, while the local galaxies
were not.  At a given luminosity, the clumps in the Kiso galaxies are typically 30
to 100 times more massive than the star forming regions in local galaxies.

Uncertainties in the various assumptions that have been made for the population
synthesis models of clump properties should not significantly affect the results
in Figure \ref{measles_mass_mag}.  We assumed solar metallicity for the local and
Kiso galaxies, and 0.4 times the solar metallicity for the UDF galaxies.  These
are reasonable assumptions considering the higher star formation rates at the same
stellar mass in UDF galaxies \citep[e.g.,][]{mannucci10}, but if the average
metallicity of the UDF galaxies were different, then the derived masses could be
different too. The effect is small, however. \cite{elmegreen09b} showed that a
decrease in metallicity from 0.008 to 0.0004 increases the mass and age by only
50\% each as a result of a decrease in line-blanketing and bluer intrinsic colors.
Uncertainties in extinction affect the age more than the mass. Higher extinction
requires bluer and younger stellar regions to give the same color, but these
younger regions have higher light-to-mass ratios which offset the greater
extinction, giving about the same required mass \citep{bell01}. As a result, the
mass uncertainties from both extinction and metallicity uncertainties are much
less than the differences in mass between the three groups of galaxies in Figure
\ref{measles_mass_mag}.

\subsubsection{The Jeans mass and implications for gas velocity dispersion}

If the star-forming clumps in all of these galaxies can be identified with the
ambient Jeans mass in the disks, and if this mass varies as $\sigma^4/\Sigma$ for
gas velocity dispersion $\sigma$ and mass column density $\Sigma$, then we can use
the results in Figure \ref{measles_mass_mag} to see how $\sigma$ and $\Sigma$ vary
from normal galaxies to Kiso clumpy galaxies to UDF clumpy galaxies.  First,
consider the low luminosity galaxies, $M_B=-16$, where the UDF clumps are more
massive than the Kiso clumps by a factor of $\sim28$ according to equation
(\ref{eq:mass}), and the UDF clump column densities are higher than the Kiso clump
column densities by a factor of $\sim1.8$ according to equation (\ref{eq:sigma}).
Resolution effects discussed above suggest that the intrinsic UDF clump column
densities are higher than the observed by an average factor of 1.4, and the KISO
column densities are higher than the intrinsic by a factor of 1.16. Thus the
intrinsic column density ratio should be corrected upward by a factor of
$1.4/1.16=1.2$. We therefore take $\Sigma_{\rm UDF}/\Sigma_{\rm Kiso}=2.2$.
Scaling these observed stellar column densities to the Jean gas column densities
(which assumes a fixed star formation efficiency), we get a ratio of velocity
dispersions from this and from the ratio of the UDF clump mass, $m_{\rm UDF}$, to
the Kiso clump mass, $m_{\rm Kiso}$,
\begin{equation}
{{\sigma_{\rm UDF}}\over{\sigma_{\rm Kiso}}}=\left({{\Sigma_{\rm UDF}}\over {\Sigma_{\rm
Kiso}}}\times{{m_{\rm UDF}}\over{m_{\rm Kiso}}}\right)^{0.25}\sim2.8
\end{equation}


Comparing the Kiso clumps to normal galaxy clumps in the same way for $M_{\rm
B}=-16$, we see that $m_{\rm Kiso}/m_{\rm normal}\sim36$ and $\Sigma_{\rm
Kiso}/\Sigma_{\rm normal}\sim1.4$ for the resolution-corrected $\Sigma_{\rm Kiso}$,
in which case $\sigma_{\rm Kiso}/\sigma_{\rm normal}\sim2.7$. Thus at low
luminosity, the mass column density of star forming material and the resulting star
complexes increase by factors of 1.4 and 2.2 going from normal galaxies to Kiso
clumpy galaxies to UDF clumpy galaxies, and the gaseous velocity dispersion
increases by factors of 2.7 and 2.8 again along this same sequence.

At high luminosity, $M_{\rm B}=-20$, the UDF and Kiso clump masses and column
densities are about the same, $m_{\rm UDF}/m_{\rm Kiso}=1.0$, and $\Sigma_{\rm
UDF}/\Sigma_{\rm Kiso}=1.3$ (corrected), so the velocity dispersions also have to
be about the same in the Jeans model, $\sigma_{\rm UDF}/\sigma_{\rm Kiso}=1.1$.
There is a big jump in clump properties from normal to Kiso clumpy galaxies,
however. Figure \ref{measles_mass_mag} and equation (\ref{eq:mass}) suggests that
$m_{\rm Kiso}/m_{\rm normal}\sim120$ at $M_{\rm B}=-20$ and $\Sigma_{\rm
Kiso}/\Sigma_{\rm normal}\sim4.0$ (corrected Kiso), which means that $\sigma_{\rm
Kiso}/\sigma_{\rm normal}\sim4.7$. Thus at high luminosity, $\Sigma$ increases by
factors of 4.0 and 1.3 going from normal to Kiso clumpy to UDF clumpy, while
$\sigma$ increases by factors of 4.7 and 1.1 in this sequence.

According to this scenario, the velocity dispersion in Kiso galaxies is higher
than in normal galaxies by factors ranging from 2.7 to 4.7, with the most luminous
Kiso galaxies having the highest excess dispersions. The column densities in Kiso
galaxies are higher than in normal galaxies by factors ranging from 1.4 to 4.0,
with the higher excess again in the most luminous Kiso galaxies. The column
densities and velocity dispersions are higher still in UDF galaxies compared to
normal galaxies, as already known \citep[see review in][]{shapley11}.  Thus the
clumpy Kiso galaxies are intermediate between normal galaxies and UDF clumpy
galaxies at low luminosity, and they are similar in clump properties to UDF
galaxies at high luminosity.

\subsubsection{Mass and Size}

Figure 5 shows the clump masses as functions of their physical sizes for all three
galaxy types. The UDF clumps are shown on the left and repeated as red plus signs
in the middle and right hand panels for comparison to the Kiso and normal galaxy
clumps. As noted in the figure caption, the Kiso and normal galaxy clumps were
measured with three different angular radii, so the masses are shown for each size
as connected lines. The sizes, given in kpc, are proportional to the galaxy
distances. The average slope of all the lines is shown as a dashed green line in
the middle and right hand panels. These slopes are 1.36 and 0.87 for Kiso and
normal galaxies, respectively.

The trend of increasing mass with clump size is not the result of background
emission because the average background has been removed by the IRAF routine {\it
apphot}. It is the result of tapered light profiles in the clumps. The average
surface density decreases by a factor of $\sim2$ from the minimum to the maximum
clump size in each survey, considering the scaling of mass with size.

The UDF clumps are all fairly large, $>0.5$ kpc, partly as a result of limited
spatial resolution at intermediate redshifts. Similarly large clumps are in the
Kiso galaxies, although their masses are slightly lower for a given size
(consistent with the lower $\Sigma$ for Kiso clumps, as discussed above). In
addition, the largest Kiso clumps have the same mass as the largest UDF clumps,
although the sizes of these largest Kiso clumps are larger than the sizes of the
largest UDF clumps (the $\Sigma$ difference again).

Star-forming complexes in normal galaxies tend to be smaller than the largest clumps
in either UDF clumpy galaxies or Kiso clumpy galaxies, but where they overlap in
size, the clumps in normal galaxies are less massive than clumps of the same size in
UDF galaxies (again this is the $\Sigma$ difference discussed above).

Clump size in the Jeans model scales with $\sigma^2/\Sigma$. If at high
luminosity, $\Sigma$ is in the sequence 1:4.0:5.1 for normal:Kiso:UDF galaxies and
$\sigma$ is in the sequence 1:4.7:5.0, as derived above, then size is in the
sequence 1:5.5:4.9, which is about what Figure 5 shows, i.e., both Kiso and UDF
galaxies have physically larger clumps than normal galaxies, by a factor of 3-10.
At low luminosity, for the above sequences of $\Sigma$ (1:1.4:3.0) and $\sigma$
(1:2.7:7.4), the Jeans size increases in the sequence 1:5.2:18.4 from normal to
Kiso to UDF galaxies.

\section{Star Formation Rates}

The average star formation rates were determined for each clump in all three samples
by taking the ratio of clump mass to age. In addition, the current star formation
rates for Kiso galaxies were determined from the H$\alpha$ equivalent widths in the
SDSS fiber spectra. These spectra have a resolution of $\sim150$ km s$^{-1}$ at
H$\alpha$ and an angular size of 3$^{\prime\prime}$ (0.16 to 3 kpc for our
distances), centered on the galaxy.  The H$\alpha$ fluxes for the full galaxies were
estimated from the H$\alpha$ equivalent widths in the fiber multiplied by the
$r$-band luminosities of the galaxies. This procedure assumes that the full galaxy
is forming stars at the rate of the central 3 arcsec. The corresponding star
formation rates were inferred from the total H$\alpha$ fluxes using the conversion
in \cite{ken}.

Figure \ref{measles_sfr_mag} shows the star formation rates for the three types of
galaxies studied here. The rates determined from the ratios of mass to age are the
dots.  The rates determined for the Kiso galaxies from the H$\alpha$ equivalent
widths in the central $3^{\prime\prime}$ extrapolated to the whole galaxy are the
red squares in the center panel. The lines show the linear fits (for these log-log
plots), with the fits for UDF and normal galaxies repeated in the central panel for
comparison, as in Figure \ref{measles_mass_mag}.  The similarity for Kiso galaxies
between the star formation rates in individual clumps and the extrapolated whole
galaxy rates suggest that the measurements in the central $3^{\prime\prime}$ fibers
are not representative of the most active clumpy regions, which are often at the
outskirts.

The star formation rates in UDF and Kiso clumps are about the same for a given
restframe absolute B magnitude.  The rates are systematically lower in the normal
galaxies by a factor of 10 to 30. This lower rate for normal galaxy clumps reflects
the lower masses of the clumps.

\section{Comparison with other studies}

The trends in Figure \ref{measles_mass_mag} and in the above discussion are
consistent with those in \cite{wisnioski12} in the sense that larger clumps in
both radius and mass also have higher velocity dispersions, with a smooth trend
going from local star-forming regions to those at $z\sim1-2$. \cite{wisnioski12}
measured velocity dispersions from H$\alpha$ emission in HII regions, while our
dispersions are assumed to represent neutral gas before gravitational
instabilities make clumps.  The high-redshift clumps in \cite{wisnioski12} have
sizes that are larger than ours by a factor of $\sim2$, and dispersions of
$50-150$ km s$^{-1}$, increasing with size to the power 0.42. Our UDF clumps have
dispersions $\sim5$ times larger than clumps in local galaxies, which converts to
$\sim50-75$ km s$^{-1}$. Thus the upper range for our dispersions is comparable to
the lower range for the \cite{wisnioski12} dispersions. This difference is not
unreasonable considering our lower clump sizes and the different phases of
interstellar gas used for the probes.

Also for \cite{wisnioski12}, the galaxies with clump sizes and masses between
those of normal local galaxies and high redshift clumpy galaxies were major
mergers \citep[from][]{monreal07}. In the present paper, they are Kiso types Ic
and Ig, chosen to be unlike mergers by their lack of obvious tidal features or
peripheral asymmetries. Still, the Kiso galaxies apparently have high gas velocity
dispersions, perhaps related to the generally asymmetric positions of their inner
red bulges and disks (Fig. 1-3).

The clumps in five $z\sim2$ galaxies studied by \cite{genzel11} are more massive
than those here because they choose regions that are bright enough to give
spectra. Their galaxies are generally brighter and more massive too: the galactic
restframe V-band AB magnitudes in their study averaged $\sim-23$ \citep{for09}
compared to our brightest B-band AB magnitude of $\sim-20$ in the region of
overlap with the Kiso sample.

\section{Radial Profiles}
The Kiso galaxies in our sample have approximately exponential light profiles.
Figure \ref{measles_ellipse2} shows ellipse-fit average radial profiles for the
Kiso galaxies. The number of scale lengths in the Kiso exponentials are given in
the figure next to each Kiso name. Some of the Kiso galaxies have a change of
slope partway out in their disks, indicating a double exponential. In most cases,
these galaxies have off-center bulges, so the exponential profile is normal until
the radius of the limit of symmetry is reached, and then the slope changes beyond
(see, e.g., Kiso 7186).

The major axis intensity profiles of Kiso galaxies are slightly more irregular than
the azimuthally averaged radial profiles because of the clumps, but the underlying
disks in Kiso galaxies are still exponential. Figure \ref{measles_pvec} shows these
major axis profiles. Some galaxies have clumps in the g-band that are as prominent
as the bulges (e.g., Kiso 4291, 4876, 5506, 5886, 5979, 6630, 8116).

Figures \ref{measles_udf_ellip} and \ref{measles_udf_pvec} show the average radial
profiles from ellipse fits and major axis profiles, respectively, for the UDF
galaxies studied here. The strips used for the major axis scans are indicated on
images of the galaxies in Figures \ref{UDF_cc15num} and \ref{UDF30num}. The
profiles were corrected for surface brightness dimming by multiplying each
intensity by $(1+z)^4$ for corresponding redshift $z$. They were not put in a
common restframe wavelength, however, because unlike the total galaxy magnitudes,
the high-resolution images of UDF galaxies have only optical coverage in the ACS
bands. Thus there is an unavoidable shift toward shorter wavelength restframes
with increasing redshift.

The average radial profiles of the UDF galaxies are roughly exponential (the number
of scale lengths is given in parenthesis), although the major axis strips are
usually irregular because of the clumps (and especially so because the restframe
wavelengths tend to be blue or UV).  Many UDF galaxies are ring-like
\citep{elmegreen09b} and their centers are hard to define for the average profiles,
but if the galaxies have bulge-like clumps near their centers then the average light
profiles away from these clumps are somewhat exponential. The average radial
positions of the clump centers are also exponential on average for the clumpy UDF
galaxies, as found in an earlier study \citep{elmegreen05b}.

\begin{figure}
\plotone{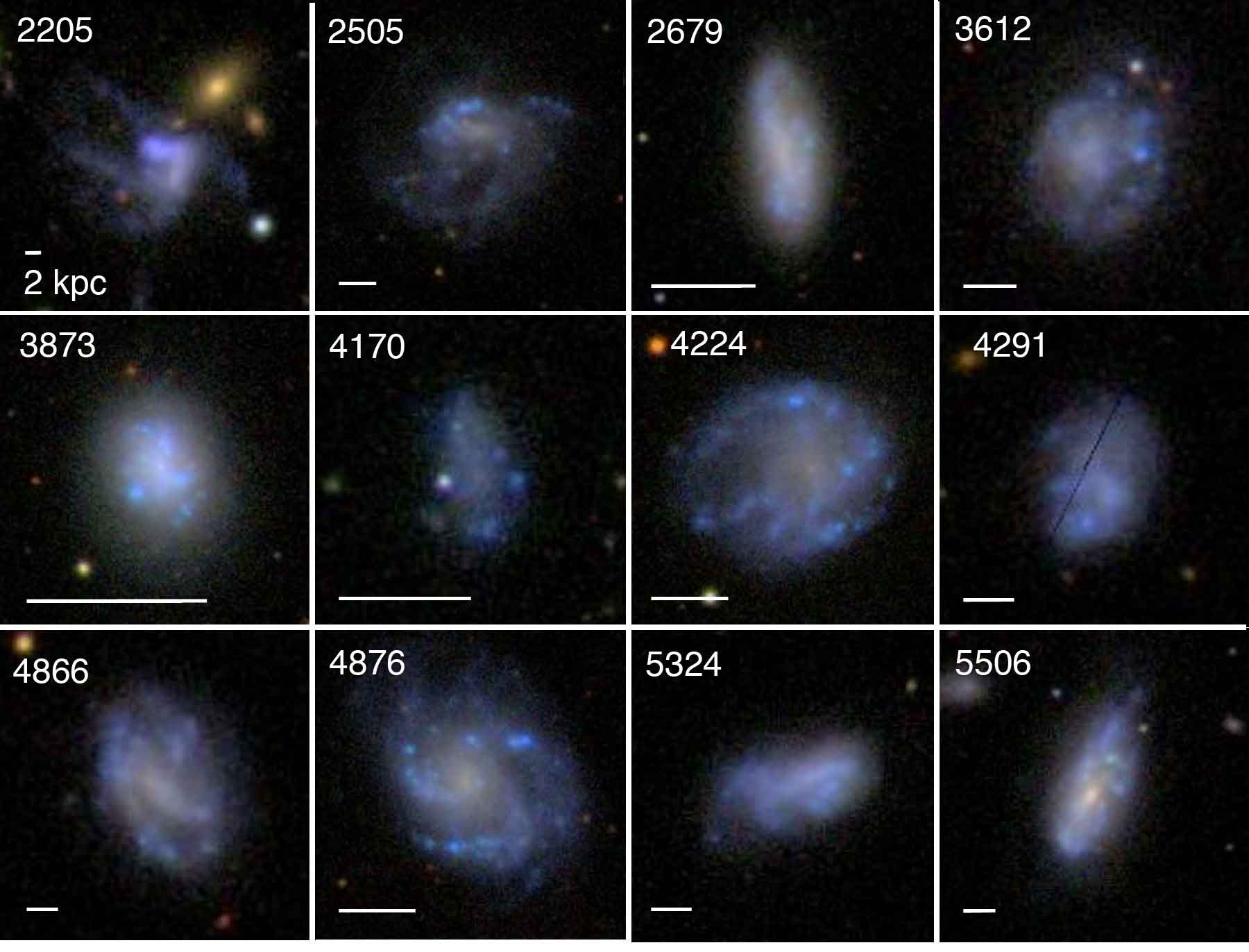}
\caption{Kiso galaxies with Kiso numbers indicated and scale bars equal to 2 kpc.
Color images are from the SDSS in g, r, and i passbands.
} \label{kisomeasles1newscale}
\end{figure}

\begin{figure}
\centering
\plotone{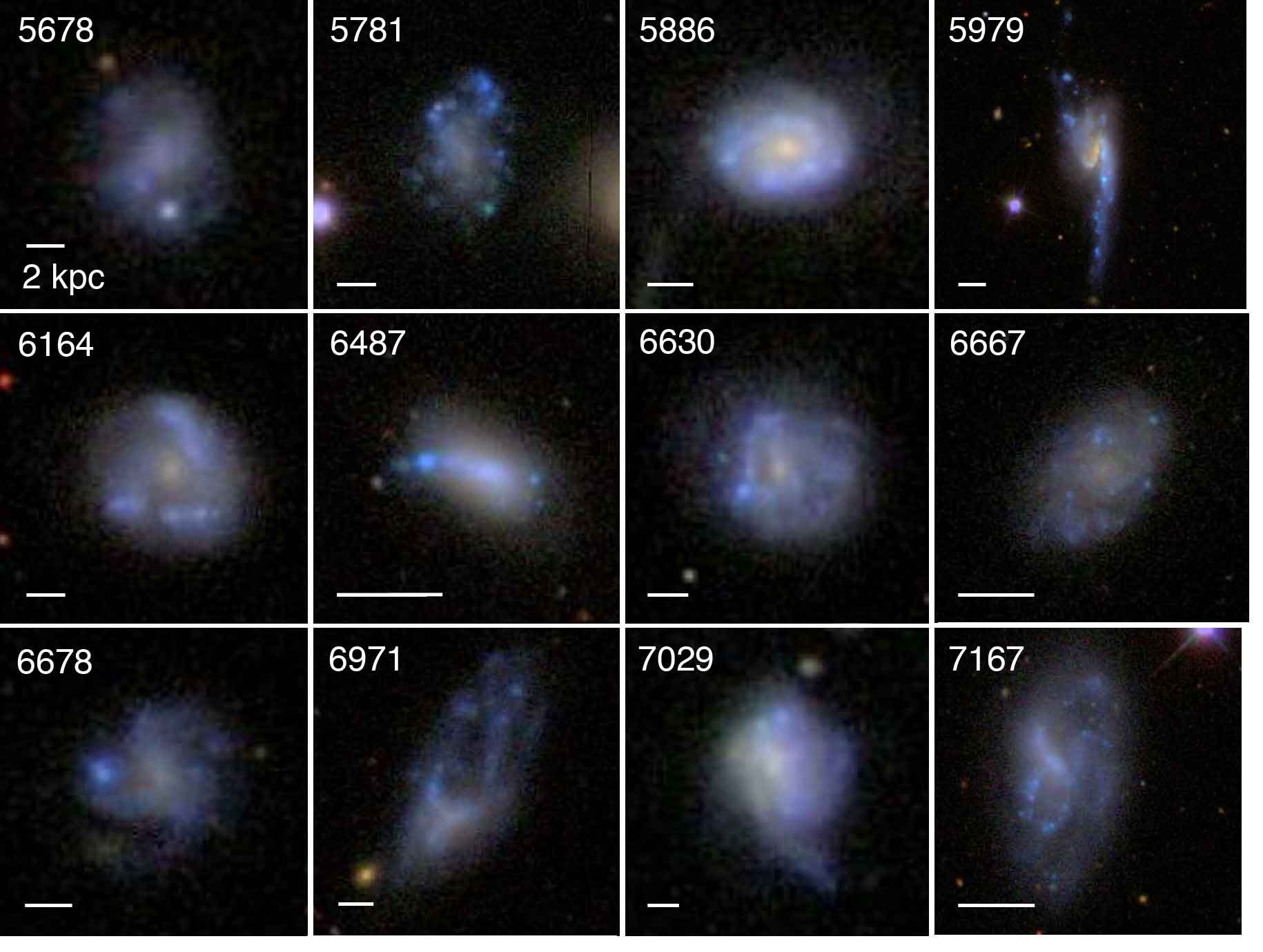}
\caption{Continues from Figure 1; Kiso galaxies with Kiso numbers indicated and scale bars equal to 2 kpc.
Color images are from the SDSS in g, r, and i passbands.
}
\label{kisomeasles2newscale}
\end{figure}

\begin{figure}
\centering
\plotone{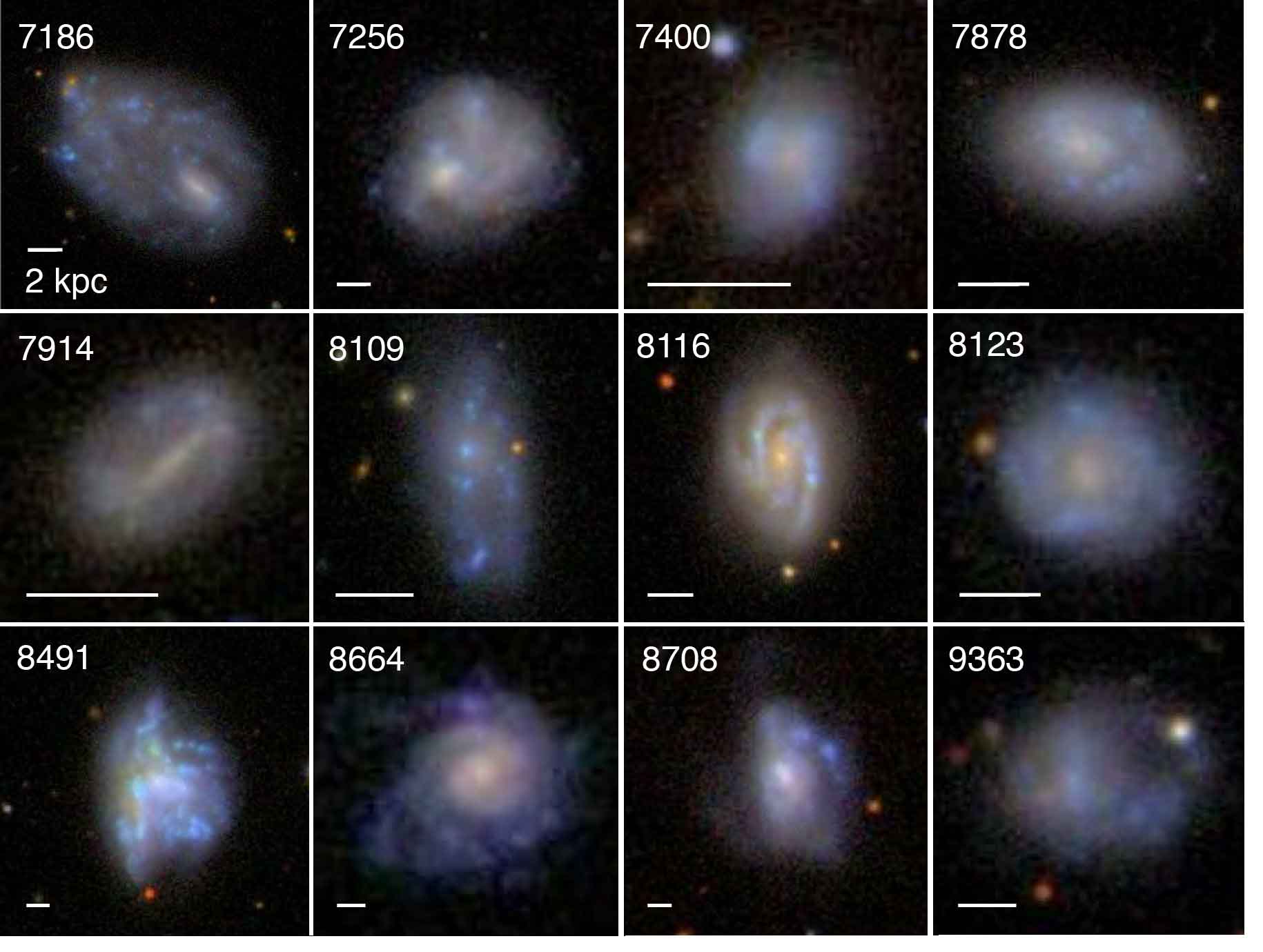}
\caption{Continues from Figure 2; Kiso galaxies with Kiso numbers indicated and scale bars equal to 2 kpc.
Color images are from the SDSS in g, r, and i passbands.
}
\label{kisomeasles3newscale}\end{figure}

\begin{figure}
\centering
\plotone{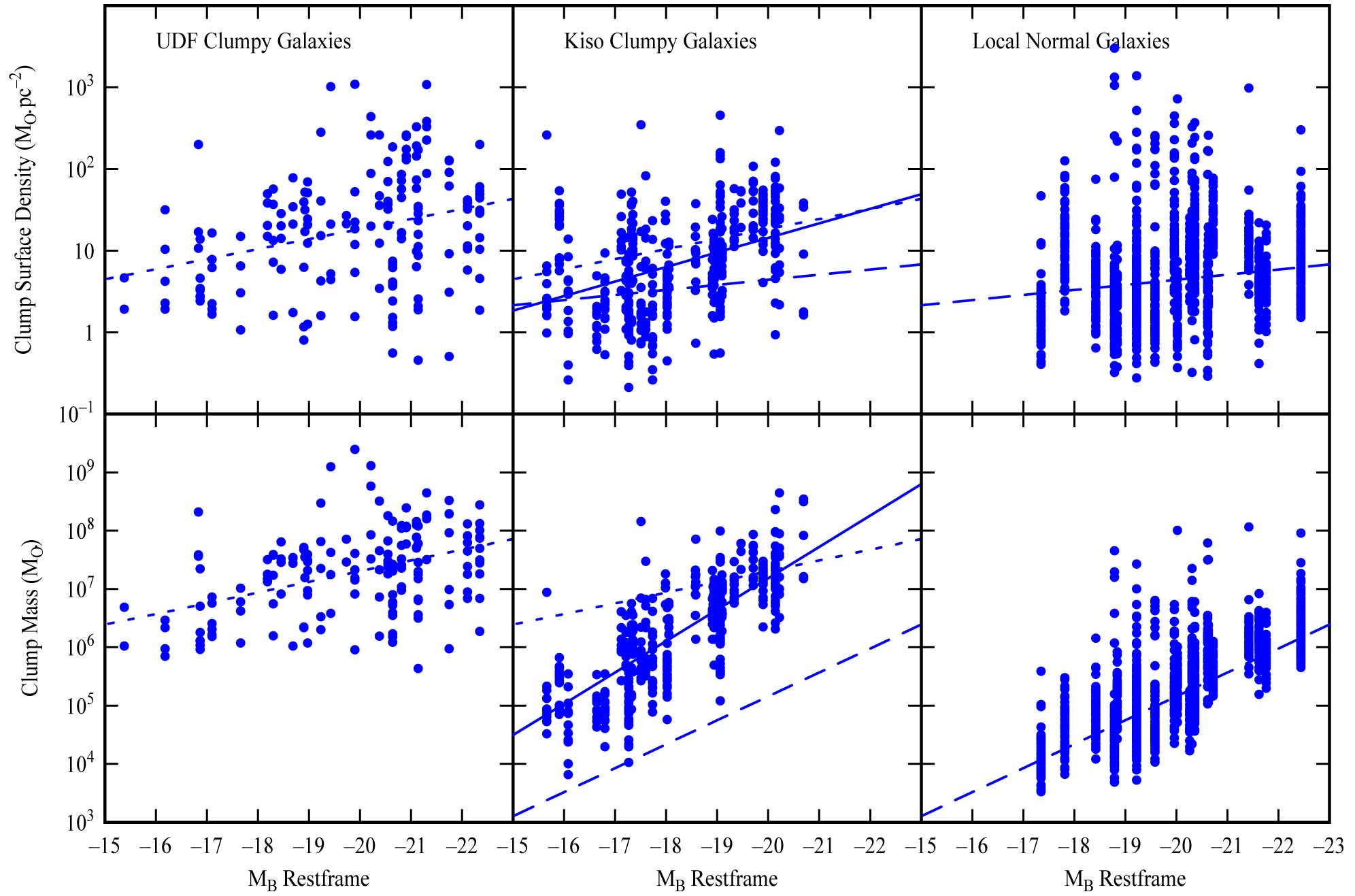}
\caption{The mass (bottom) and surface density (top) of clumps in UDF (left), Kiso (middle),
and normal spiral galaxies (right), plotted versus the absolute restframe B magnitude of the
whole galaxy.
Fits to the data are shown in each panel, with the fits in the central panels shown by solid lines and
the fits to the UDF and normal clumps repeated in the central panels for comparison.
}\label{measles_mass_mag}\end{figure}

Figures \ref{normalspirals} to \ref{threenormals} show three normal spirals, a grand
design (NGC 5248), multiple arm (NGC 3344) and a flocculent (NGC 5055), for
comparison.  Figure \ref{normalspirals} has SDSS sky images with the same color
scheme as in Figures \ref{kisomeasles1newscale} - \ref{kisomeasles3newscale}, Figure
\ref{measles_pvec_normal} shows the ellipse fit radial profiles on the left and
major axis scans on the right, and Figure \ref{threenormals} shows the scan
directions for the major axis profiles.   The normal galaxies differ from the Kiso
and UDF galaxies because the normals have a more prominent bulge and a smoother
radial profile than the other types (the sharp peaks in the ellipse fits for NGC
3344 and NGC 5248 are foreground stars). This smoothness is consistent with the
previous result that the non-bulge clumps in normal galaxies are smaller than those
in Kiso and UDF galaxies.

\section{Discussion}
We find that the brightest Kiso galaxies have giant non-bulge clumps that are
similar in mass and surface density to the non-bulge clumps in high-redshift
clumpy galaxies of the same absolute restframe B magnitude. The local normal
galaxies have much smaller clumps than either of these. This trend in clump mass
and surface density is consistent with a steady decrease over time in average gas
surface density and turbulent speed in disk galaxies, leading to a corresponding
decrease in the Jeans mass.

The high mass clumps in the Kiso galaxies are often in spiral arms, but analogous
arms cannot be seen in the clumpy UDF galaxies. Part of the difference in UDF
structure could be from the lack of sufficiently deep, high resolution restframe
V-band or redder observations, where local galaxies show spiral structure in the
old disk stars. However, the restframe uv sampled in the clumpy UDF galaxies does
not show spiral features either, whereas GALEX uv images of local galaxies do show
spiral arms, even if only in the alignment of clumps. In addition, clumpy galaxies
without spirals were found in lower-redshift surveys too (Elmegreen et al. 2009b),
and their restframe B or V-band images have sufficiently high resolution and depth
in the ACS camera to have shown spirals if they were present.

Kiso galaxies are also midway between UDF and normal galaxies with respect to the
average radial disk profiles. While all three types have roughly exponential
profiles for azimuthally-averaged light when measured with increasing distance from
the brightest clumps or bulge-like centers, UDF galaxies viewed in restframe blue or
uv light have the most irregularity in these profiles, with Kiso galaxies next. Even
more extreme are major axis intensity scans, which are highly irregular in the UDF
clumpy galaxies and still irregular but less so in the Kiso galaxies.

Galaxy evolution is characterized by the simultaneous build up of an old stellar
disk and the loss of giant star-forming clumps. Such clumps are still possible in
the local universe in exceptional circumstances, presumably when the gas becomes
highly turbulent. The giant clumps in the Kiso galaxies, for example, are mostly in
spiral arms, where even local galaxies have somewhat massive clumps, whereas high
redshift clumpy galaxies do not have spirals.  The source of the presumed high
turbulence in clumpy Kiso galaxies could be related the overall disk asymmetry.

\section{Conclusions}

Star forming complexes and radial disk profiles were studied in three samples of
clumpy irregular galaxies, one at high redshift using the Hubble UDF, and two at low
redshift using UV-bright Kiso galaxies and normal spirals. The brightest Kiso
galaxies have clump masses and surface densities comparable to those at high
redshift for the same absolute restframe B-band magnitude. These two types also have
irregular radial profiles because of the clumps, although the UDF galaxies are more
irregular. Normal local spirals have smoother profiles and clumps that are lower in
mass by a factor of $\sim100$. Clump ages in the Kiso and local spiral galaxies are
on the order of 10 Myr, with slightly older ages for the UDF clumps.

If the clumps have masses comparable to the local Jeans' mass, then this mass is
related to the clump surface density and velocity dispersion. Using this relation,
we determine that the interstellar velocity dispersion increases from local spirals
to Kiso clumpies to UDF clumpies in the proportion 1 to 4.7 to 5.0 for the most
luminous galaxies in our sample, and in the proportion 1 to 2.7 to 7.4 for the least
luminous, again for a given restframe absolute magnitude.  The surface densities
also increase along this sequence from 1 to 4.0 to 5.1 for the most luminous, and 1
to 1.4 to 3.0 for the least. We therefore predict that many of the clumpy Kiso types
Ic and Ig will have elevated velocity dispersions and gas surface densities compared
to local spiral galaxies. Considering both their clump properties and the radial
profiles, Kiso galaxies are intermediate between highly clumpy UDF galaxies and
local spirals.

Kiso galaxies are massive disk galaxies like local spirals, based on their absolute
magnitudes, but they contain bigger and denser clumps than the spirals, and this
makes them resemble young galaxies in the UDF.  There is no reason to think that
clumpy Kiso galaxies are young, so this morphological property can evidently be
triggered in the modern universe by increasing the surface density and velocity
dispersion in a galaxy.  Such increases might result from a near encounter with
another galaxy of comparable mass, or from other types of torquing, which stir the
disk and lead to accretion. We note that most of the Kiso galaxies in our survey
have bulges that are slightly offset from the center, and this asymmetry may drive
turbulence and disk inflow also.

\clearpage

\begin{deluxetable}{lclcccc}
\tabletypesize{\scriptsize}  \tablecaption{Kiso Galaxies} \tablehead{ \colhead{KUG}
& \colhead{Kiso} & \colhead{Other} & \colhead{Type} &\colhead{Features} &
\colhead{Distance}
& \colhead{$M_B$}\\
&&&&&(Mpc)&(mag) } \startdata
KUG 0726+400  &   2205 &                                             & Ic &  streamer&205.6 &   -20.14\\
KUG 0757+373  &   2505 & UGC 04157                                   & Ic &off-center&   53.9 &   -17.73\\
KUG 0813+243  &   2679 & IC2556                                      & Ic &&   29.3 &   -17.35\\
KUG 0926+455  &   3612 &                                             & Ic &spiral&   59.1 &   -17.60\\
KUG 0937+485  &   3873 & Mrk 1418, UGC 05151                         & Ic &&   11.0 &   -15.91\\
KUG 0950+295  &   4170 &                                             & Ic& &   21.6 &   -14.10\\
KUG 0952+335  &   4224 & UGC 05326                                   & Ic&off-center &   19.0 &   -16.80\\
KUG 0956+439B &   4291 &                                             & Ic &&   69.9 &   -17.23\\
KUG 1045+334  &   4866 &                                             & Ic &off-center&   91.6 &   -18.91\\
KUG 1046+330  &   4876 & IC 2604, UGC 05927, VV 538                  & Ic&spiral &   22.1 &   -17.27\\
KUG 1125+240  &   5324 &                                             & Ig&bar &   80.5 &   -19.09\\
KUG 1131+159  &   5506 &                                             & Ig &streamer&   70.2 &   -19.33\\
KUG 1140+331  &   5678 &                                             & Ic &&  135.4
&   -18.58\\  Irr
KUG 1145+130  &   5781 & near IC 0737                                & Ic& &   48.8 &   -17.21\\
KUG 1150+353  &   5886 &                                             & Ic &spiral&   87.6 &   -19.71\\
KUG 1155+281  &   5979 & NGC 4004, VV 230, Mrk 0432       & Ic&streamer &   45.9 &   -19.06\\
KUG 1204+248  &   6164 &                                             & Ic&spiral &  105.7 &   -19.10\\
KUG 1218+179  &   6487 &                                             & Ig &BCD&   28.3 &   -17.12\\
KUG 1226+206  &   6630 &                                             & Ic
&off-center spiral&   94.5 &   -18.94\\ I
KUG 1229+207  &   6667 &                                             & Ic &spiral&   26.0 &   -16.64\\
KUG 1229+250  &   6678 & IC 3472                                     & Ic&spiral &   94.9 &   -18.02\\
KUG 1251+363  &   6971 &                                             & Ic &streamer&   58.0 &   -17.51\\
KUG 1256+274  &   7029 & Mrk 0057                                    & Ic&streamer &  105.1 &   -19.47\\
KUG 1308+368  &   7167 & NGC 5002, UGC 08254                         & Ic &bar&   15.6 &   -16.08\\
KUG 1310+230  &   7186 & NGC5012A                                  & Ic &off-center,bar,spiral&   35.4 &   -18.02\\
KUG 1319+356  &   7256 &                                             & Ic &off-center &   80.7 &   -18.92\\
KUG 1336+312  &   7400 &                                             & Ic &spiral?&   65.9 &   -17.99\\
KUG 1408+163  &   7878 &                                             & Ic &off-center&   32.3 &   -17.32\\
KUG 1409+110  &   7914 &                                             & Ic &bar&  102.2 &   -18.57\\
KUG 1425+137  &   8109 & UGC 09273                                   & Ic& &   18.0 &   -15.66\\
KUG 1426+234  &   8116 & NGC 5637                                    & Ic&spiral &   72.6 &   -19.90\\
KUG 1427+262  &   8123 &                                             & Ic &spiral?&   67.2 &   -18.05\\
KUG 1603+206  &   8491 & Mrk0297, NGC 6052, Arp 209, VV 0862 & Ic& &   66.3 &   -20.14\\
KUG 1619+408  &   8664 &                                             & Ic &spiral&  297.5 &   -20.69\\
KUG 1624+404  &   8708 & Mrk 881                                     & Ic& &  120.5 &   -20.22\\
KUG 2316+154  &   9363 &                                             & Ic
&off-center&   63.2 &   -17.51

\enddata
\label{results1}
\end{deluxetable}

\begin{deluxetable}{lcccc}
\tabletypesize{\scriptsize}  \tablecaption{SDSS Local Spiral Galaxies} \tablehead{
\colhead{NGC} &\colhead{Type} &\colhead{Arm class} & \colhead{Distance}
& \colhead{$M_B$}\\
&&&(Mpc)&(mag) } \startdata

  1087 &SAB(rs)c   & flocculent&20.80 &   -20.63\\
  2903 &SAB(rs)bc &grand design&     6.52 &   -19.96\\
  3184 &SAB(rs)cd &multiple arm&     8.17 &   -19.22\\
  3344 &(R)SAB(r)bc &multiple arm &    7.21 &   -18.79\\
  3351 &SB(r)b &grand design&    9.30 &   -19.58\\
  3433 &SA(s)c &grand design &   35.80 &   -20.61\\
  3596 &SAB(rs)c& flocculent&    15.20 &   -19.22\\
  4254 &SA(s)c &multiple arm &   32.30 &   -22.44\\
  4303 & SAB(rs)bc &multiple arm&   20.30 &   -21.42\\
  4449 & IBm&-&    3.55 &   -17.81\\
  4519 &SB(rs)d& flocculent&    15.80 &   -18.84\\
  4535 &SAB(s)c &grand design&    26.00 &   -21.76\\
  4618 &SB(rs)m &flocculent&     8.12 &   -18.42\\
  5055 &  SA(rs)bc& flocculent&   7.54 &   -20.36\\
  5194 &SA(s)bc &grand design&     7.54 &   -20.72\\
  5248 &  SAB(rs)bc &grand design&  15.40 &   -20.31\\
  5371 & SAB(rs)bc &grand design&   36.20 &   -21.62\\
  5457 &SAB(rs)cd& multiple arm&4.94 &   -20.26\\
  5474 & SA(s)cd pec &flocculent&    5.36 &   -17.35\\
  5584 &SAB(rs)cd &multiple arm&    22.10 &   -20.02
\enddata
\label{results}
\end{deluxetable}

\begin{deluxetable}{lcc}
\tabletypesize{\scriptsize}  \tablecaption{UDF Clumpy Galaxies} \tablehead{
\colhead{UDF} & \colhead{redshift}
& \colhead{$M_B$}\\
&&(mag) } \startdata
   533 &   2.42 &  -19.43\\
  1666 &   1.35 &  -20.64\\
  1681 &   0.12 &  -12.54\\
  1775 &   2.15 &  -18.97\\
  2012 &   2.98 &  -22.10\\
  2340 &   1.32 &  -18.30\\
  2350 &   3.09 &  -19.73\\
  2499 &   0.69 &  -15.38\\
  2538 &   2.56 &  -18.92\\
  3460 &   1.46 &  -16.83\\
  3483 &   2.24 &  -22.35\\
  3752 &   2.17 &  -21.14\\
  3799 &   1.46 &  -19.23\\
  3844 &   2.60 &  -21.75\\
  3881 &   0.01 &   -8.20\\
  4616 &   1.47 &  -20.22\\
  4860 &   3.55 &  -20.38\\
  4999 &   1.36 &  -18.90\\
  5107 &   3.52 &  -18.69\\
  5190 &   1.35 &  -20.81\\
  5501 &   0.01 &   -8.70\\
  5620 &   0.21 &  -16.18\\
  5685 &   0.55 &  -17.10\\
  5748 &   1.97 &  -19.90\\
  5827 &   1.66 &  -21.11\\
  5837 &   2.28 &  -18.19\\
  5878 &   2.50 &  -21.30\\
  6056 &   0.67 &  -16.87\\
  6396 &   2.64 &  -18.45\\
  6438 &   2.60 &  -20.91\\
  6821 &   1.12 &  -20.55\\
  7905 &   0.71 &  -17.66
\enddata
\label{results}
\end{deluxetable}

\clearpage

\clearpage

\clearpage

\clearpage

\clearpage
\begin{figure}
\centering
\plotone{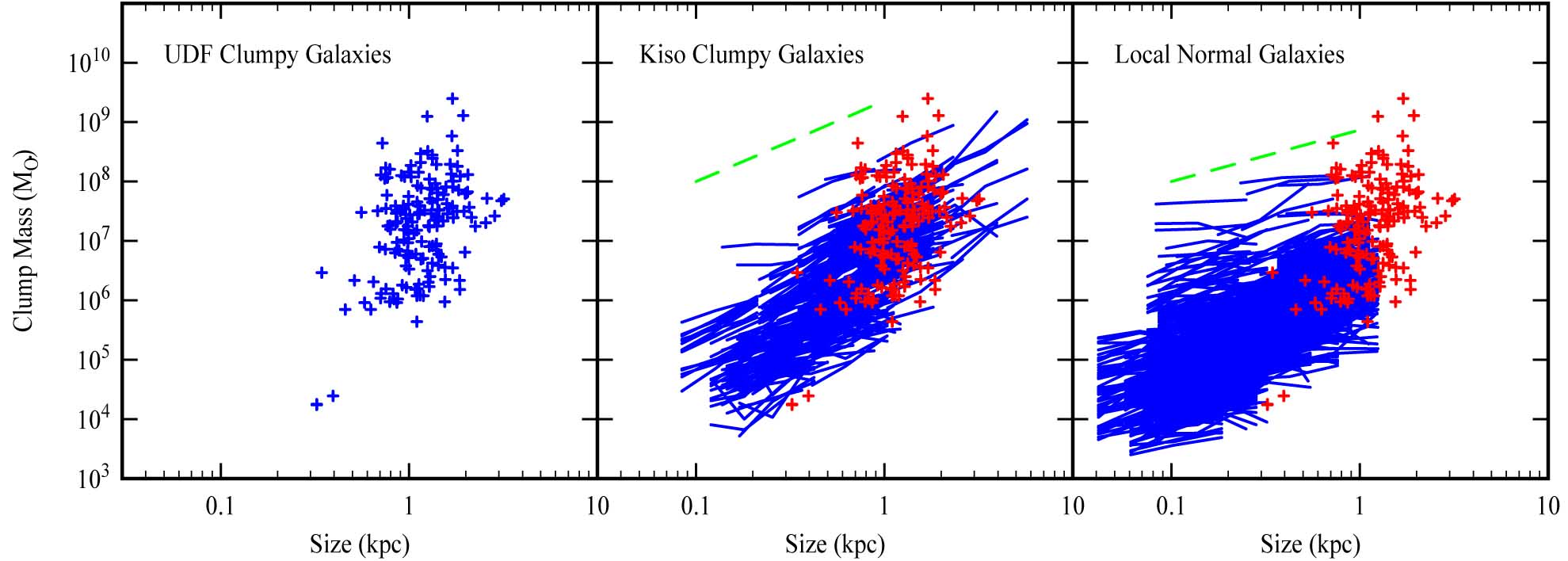}
\caption{The clump masses are shown as functions of the clump sizes (diameters) for the
three galaxy types. For Kiso and normal galaxies, the clumps were measured at three
different radii: 2, 3, and 5 pixels for Kiso galaxies, and 3, 5, and 9 pixels for
normal spiral galaxies. The lines in each figure connect these three measurements for each clump;
the green dashed lines show the average slopes of mass versus size (1.36 for Kiso galaxies
and 0.87 for local galaxies).
The UDF mass-size distribution is shown in the center and right panels as
a collection of red plus-marks for comparison.
}\label{measles_mass_size}\end{figure}

\clearpage
\begin{figure}
\centering
\plotone{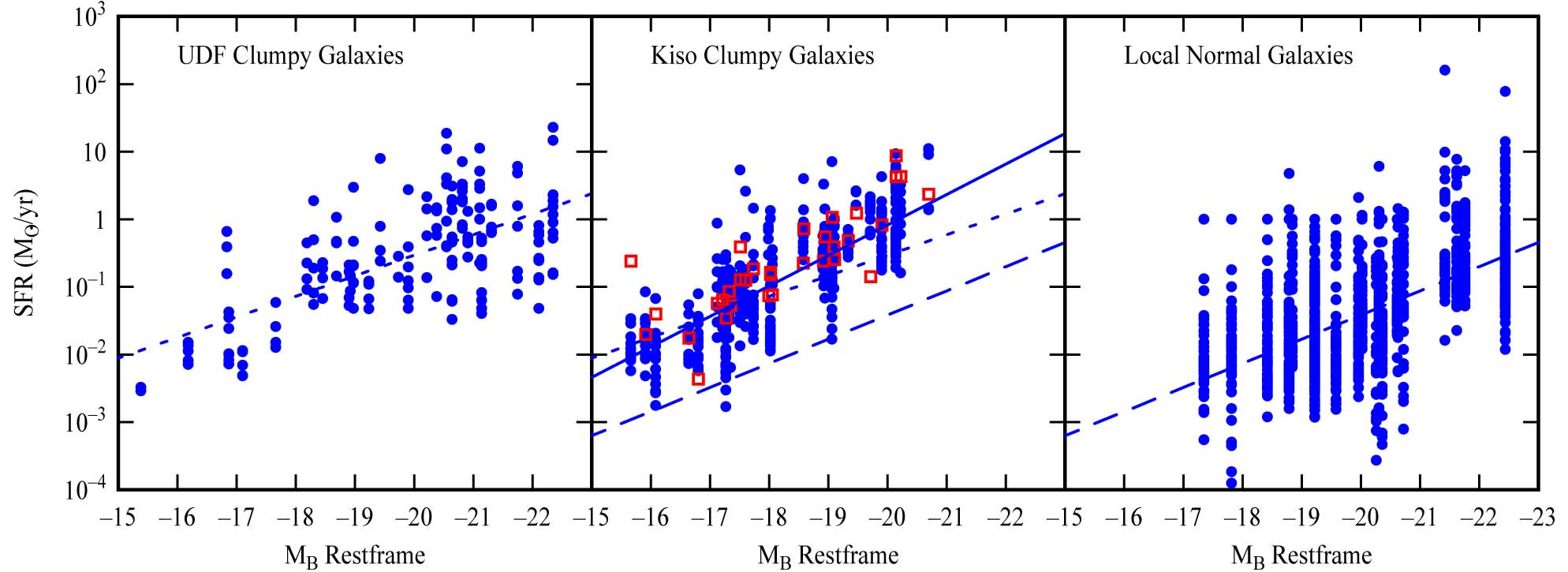}
\caption{The clump star formation rates in $M_\odot$ yr$^{-1}$ are shown versus
the absolute restframe B magnitudes of the whole galaxies for all three galaxy types.  The fits to the points
are indicted, with the UDF and normal galaxy fits repeated in the central panel for
comparison.  The red squares in the central panel are star formation rates for the whole
Kiso galaxies determined from the product of the H$\alpha$ equivalent width in the SDSS
fiber and the SDSS r-band luminosity of the galaxy.
}\label{measles_sfr_mag}\end{figure}

\clearpage
\begin{figure}
\centering
\plotone{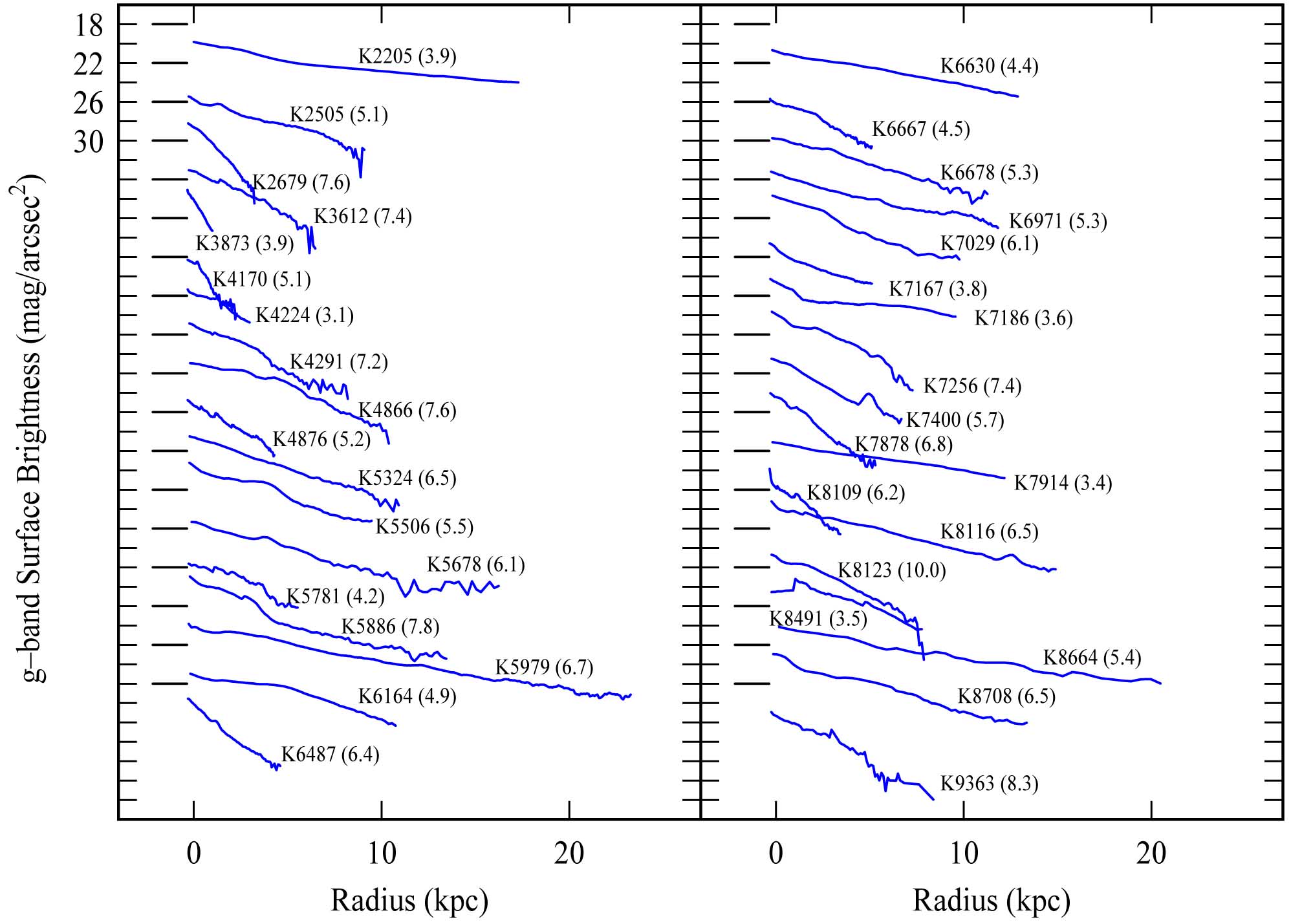}
\caption{The radial profiles of azimuthally-averaged light in g-band are shown for all Kiso
galaxies studied here. The profiles are approximately exponential but there are sometimes large
local excesses at the radii of the clumps. The Kiso numbers are indicated with
the number of exponential scale lengths in the whole profile given in parentheses.
The vertical axis ticmarks are separated by 2 magnitudes per square arcsec of
surface brightness.  Each scan has a different vertical shift in this plot. The ticmarks
on the left in each panel indicate a surface brightness of 18 mag arcsec$^{-2}$ for
the corresponding scan, which usually comes close to touching it.
The numerical labels
at the top left correspond to the surface brightness scale of the top
profile, i.e., for Kiso 2205.
}\label{measles_ellipse2}\end{figure}

\clearpage
\begin{figure}
\centering
\plotone{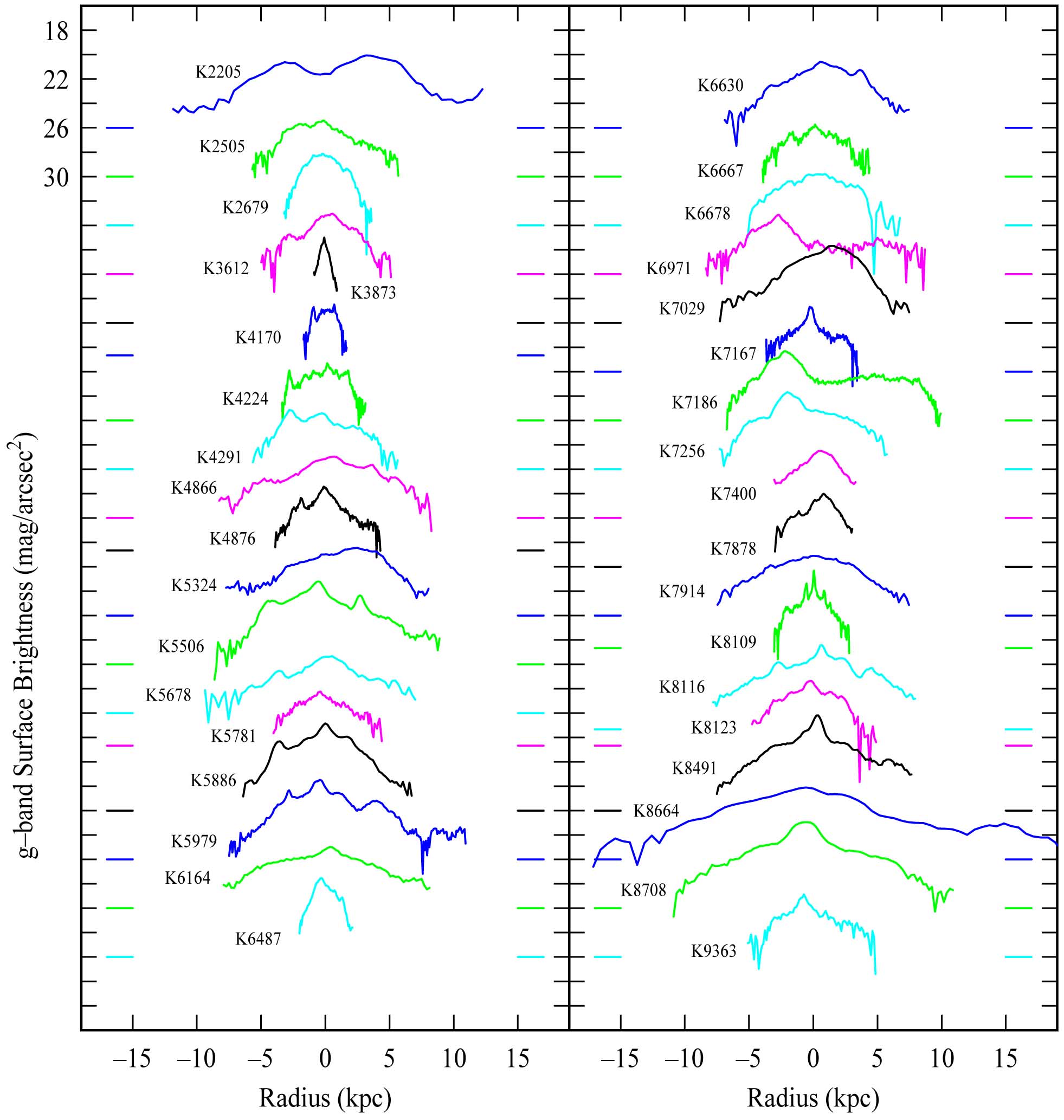}
\caption{Major axis scans through the Kiso galaxies in g-band are shown versus
position in kpc, with zero position at the center of the scan. The clumps are more
prominent in these single-direction scans than in the average radial profiles in
the previous figure. Still, most Kiso galaxies have a smooth component to the disk
with an approximately exponential profile. The vertical axis ticmarks are
separated by 2 magnitudes per square arcsec of surface brightness.  Each scan has
a different vertical shift. The colored ticmarks on the left and right of each
panel indicate a surface brightness of 26 mag arcsec$^{-2}$ for the corresponding
scan, which has the same color. Sometimes the scans are shifted up and down
slightly so they do not overlap. The numerical labels at the top left correspond
to the surface brightness scale of the top profile, i.e., for Kiso 2205.
}\label{measles_pvec}\end{figure}

\clearpage
\begin{figure}
\centering
\plotone{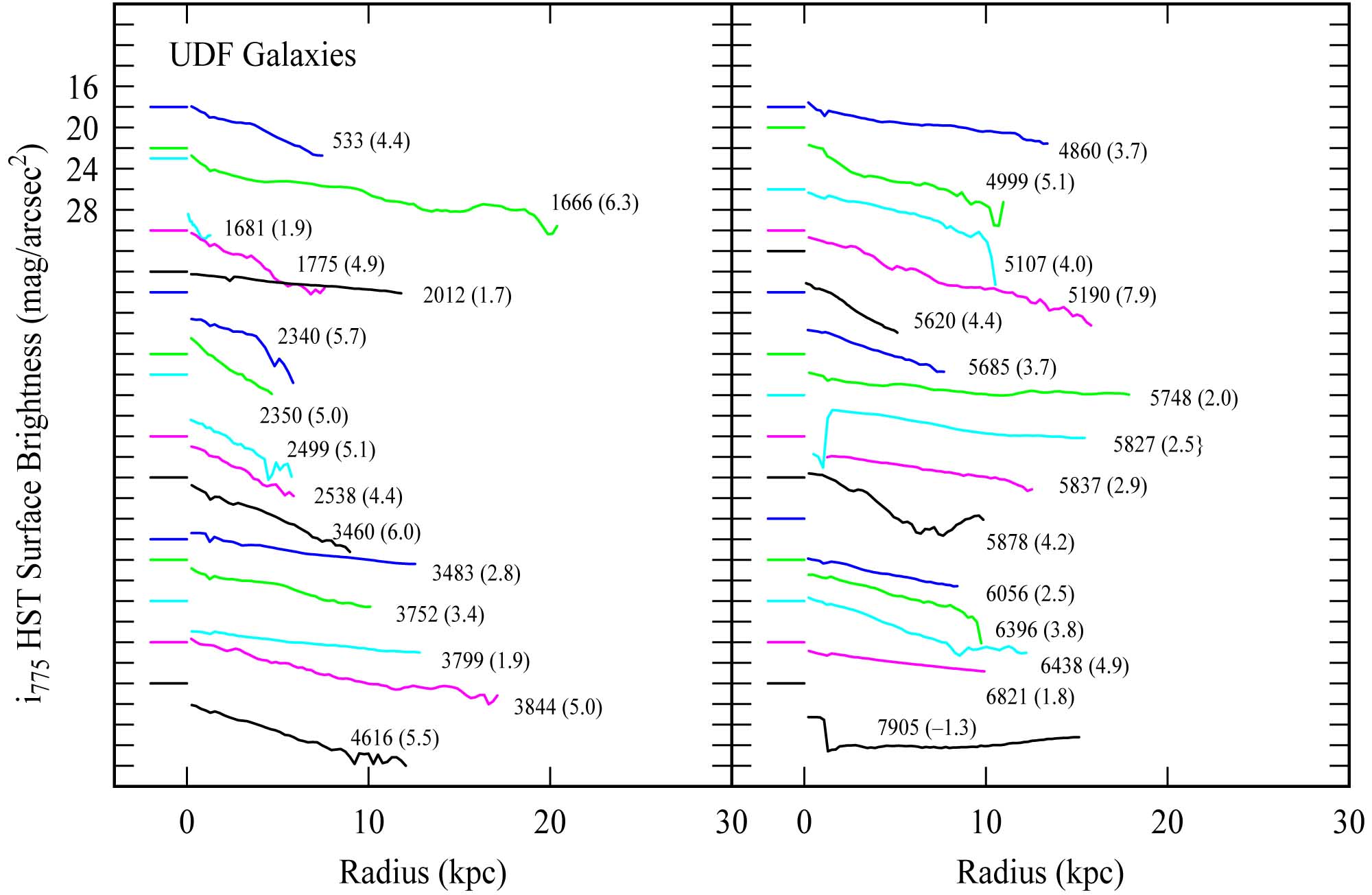}
\caption{The radial profiles of azimuthally-averaged light in HST $i_{775}$-band are shown for all UDF
galaxies studied here. The profiles are sometimes exponential but there are
usually large local excesses at the radii of the clumps. The UDF numbers are
indicated with the number of exponential scale lengths in the whole profile given in
parentheses. The vertical axis ticmarks are separated by 2 magnitudes per square
arcsec of surface brightness.  Each scan has a different vertical shift.
The ticmarks on the left in each panel indicate a surface brightness of 18 mag
arcsec$^{-2}$ for the corresponding scan, which has the same color and
usually comes close to touching it.
Sometimes the scans are shifted up and down slightly so they do not overlap. The
numerical labels at the top left correspond to the surface brightness scale of the
top profile, i.e., for UDF 533.
For the determination of the number of scale lengths indicated in parentheses, we
ignored the rapidly changing inner parts in galaxies 5827 and 7905, and the
rapidly changing outer parts in 5107 and 6396. The scans have been corrected for
surface brightness dimming but not interpolated to a common restframe wavelength.}
\label{measles_udf_ellip}\end{figure}

\clearpage
\begin{figure}
\centering
\plotone{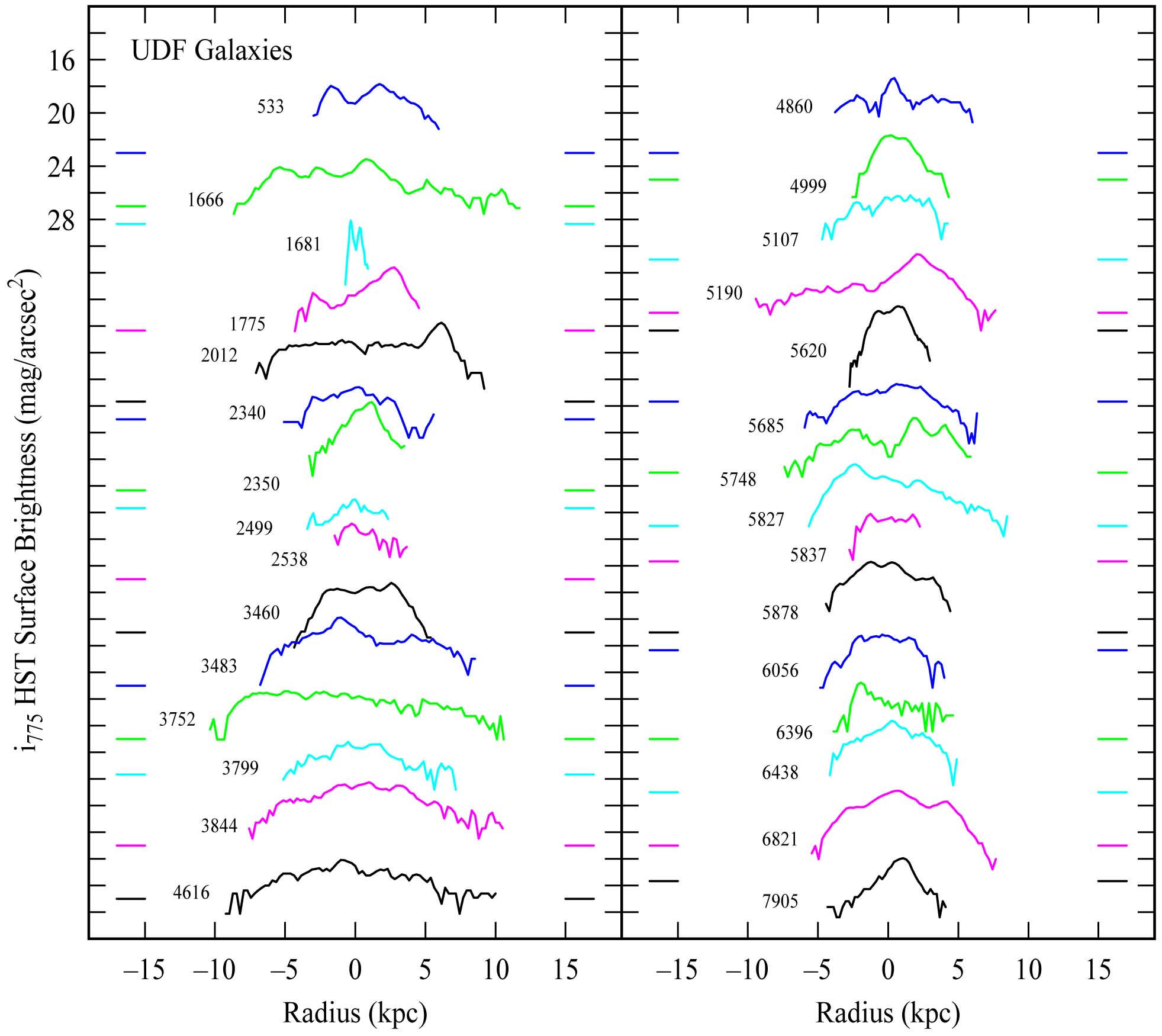}
\caption{Linear scans through the UDF galaxies (with scan directions shown in the two following
figures) in HST $i_{775}$-band are shown versus position
in kpc, with zero position at the center of the scan. The clumps are more prominent in these single-direction
scans than in the average radial profiles in the previous figure. Many UDF galaxies
have no semblance of a smooth underlying component and no exponential profile.
The vertical axis ticmarks are
separated by 2 magnitudes per square arcsec of surface brightness.  Each scan has a
different vertical shift. The colored ticmarks on the left and right of
each panel indicate a surface brightness of 23 mag arcsec$^{-2}$ for the
corresponding scan, which has the same color. Sometimes the scans are shifted up and
down slightly from a more regular positioning so that they do not overlap.
The numerical labels
at the top left correspond to the surface brightness scale of the top
profile, i.e., for UDF 533. The scans have been corrected for
surface brightness dimming but not interpolated to a common restframe wavelength.}
\label{measles_udf_pvec}\end{figure}

\clearpage
\begin{figure}
\centering
\plotone{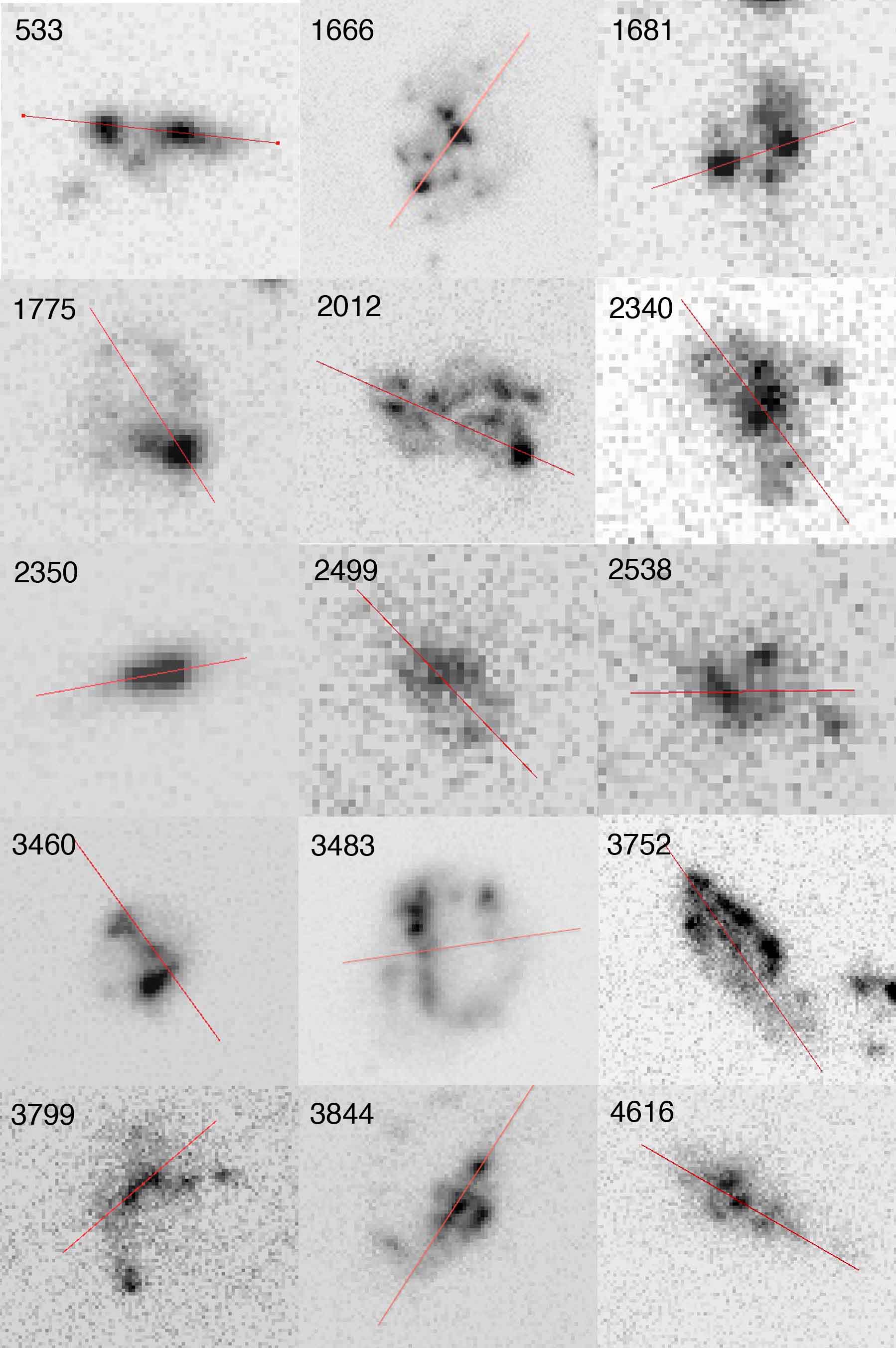}
\caption{Images of the UDF clumpy galaxies studied here with lines showing the positions
of the linear scans used for figure 10.  The passband is $i_{775}$.
}\label{UDF_cc15num}\end{figure}

\clearpage
\begin{figure}
\centering
\plotone{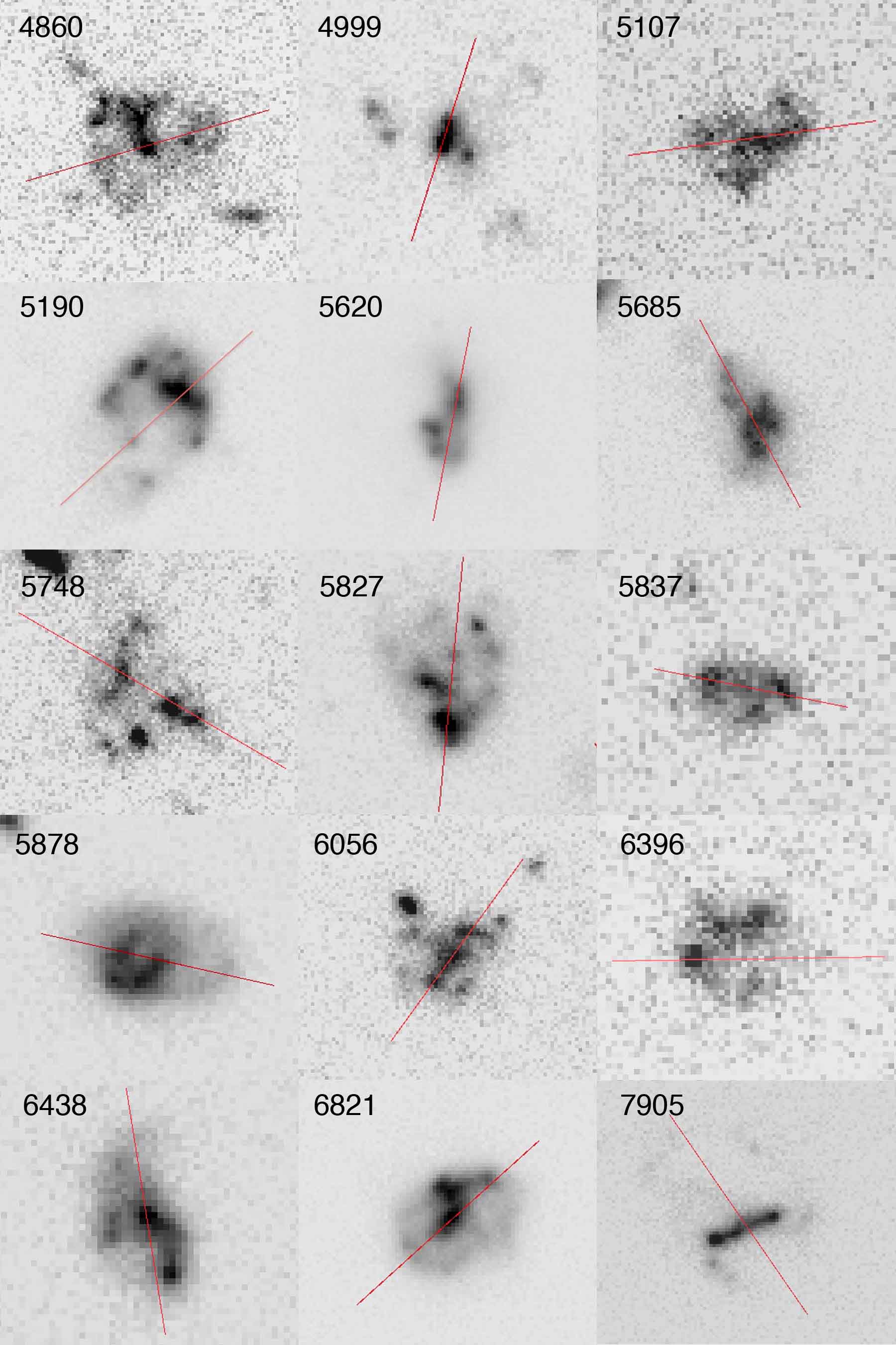}
\caption{Images of the UDF clumpy galaxies studied here with lines showing the positions
of the linear scans used for figure 10. The passband is $i_{775}$.
}\label{UDF30num}\end{figure}

\clearpage
\begin{figure}
\centering
\plotone{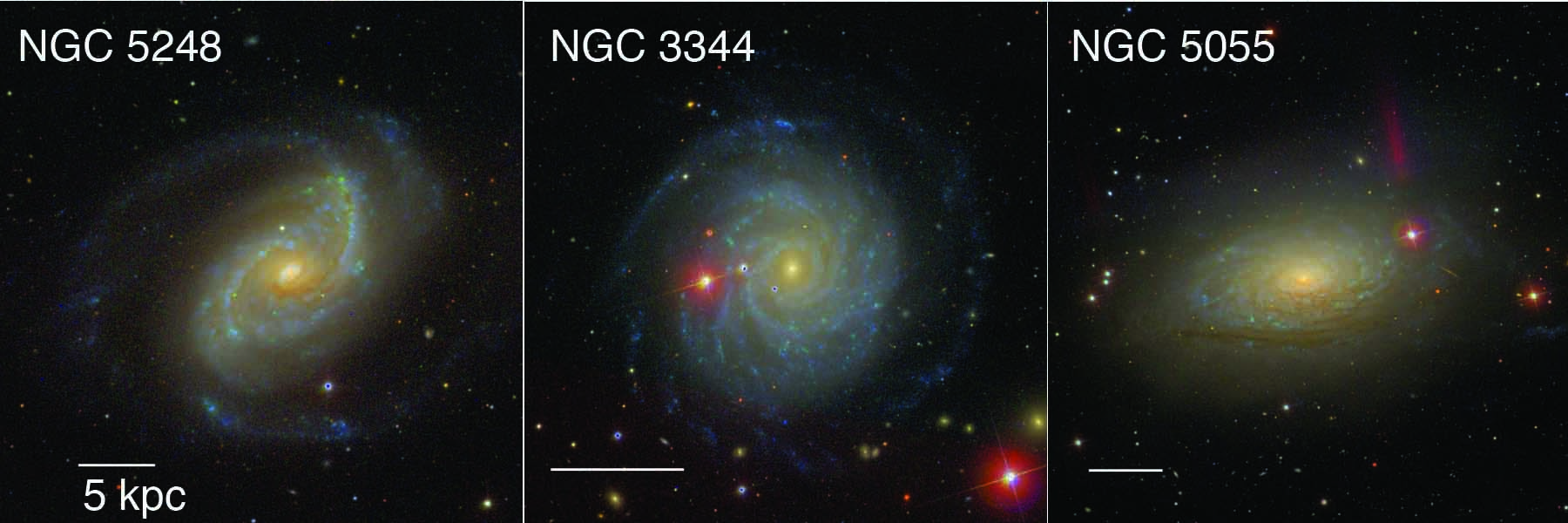}
\caption{SDSS images in g, r, and i passbands of three local galaxies that were
used to compare the radial profiles with those
of clumpy Kiso and UDF galaxies.
}\label{normalspirals}\end{figure}

\clearpage
\begin{figure}
\centering
\plotone{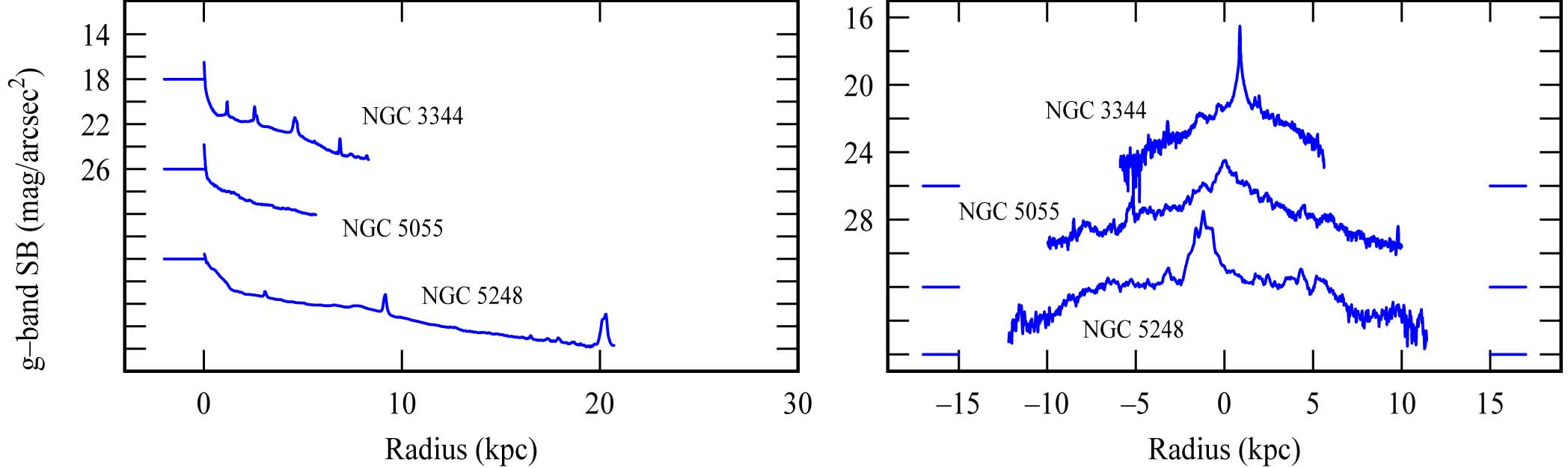}
\caption{Ellipse-fit radial profiles (left) and major axis scans (right) for
3 local spiral galaxies, in g-band. Each profile is shifted vertically
in this plot for clarity, with the y-axis numbers corresponding to the top profile.
The horizontal marks at negative radii are at surface brightnesses of 18 mag arcsec$^{-1}$
on the left, and 26 mag arcsec$^{-1}$ on the right.  The major axis profiles are centered
on the midpoints of the scans, not the galaxy centers.
}\label{measles_pvec_normal}\end{figure}

\clearpage
\begin{figure}
\centering
\plotone{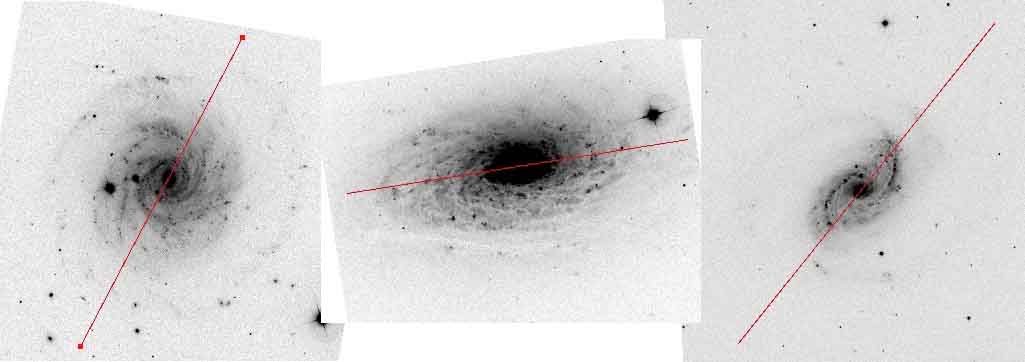}
\caption{Scan directions for the local spiral intensity profiles shown in the
previous figure.  The images are g-band from the SDSS.
}\label{threenormals}\end{figure}

\end{document}